# Informational Way to Protein Alphabet: Entropic Classification of Amino Acids

Alexander Gorban[1,2]*, Mikhail Kudryashev[3], Tatiana Popova[2]

[1]Centre for Mathematical Modeling, University of Leicester, UK, E-mail: ag153@le.ac.uk,
[2]Institute of Computational Modeling, Akademgorodok, Krasnoyarsk, Russia;
[3]Krasnoyarsk State University, Krasnoyarsk, Russia.

**ABSTRACT**

**Motivation:** What are proteins made from, as the working parts of the living cells protein machines? To answer this question, we need a technology to disassemble proteins onto elementary functional details and to prepare lumped description of such details. This lumped description might have a multiple material realization (in amino acids). Our hypothesis is that informational approach to this problem is possible. We propose a way of hierarchical classification that makes the primary structure of protein maximally non-random. The first steps of the suggested research program are realized: the method and the analysis of optimal informational protein binary alphabet. The general method is used to answer several specific questions, for example:

- Is there a syntactic difference between Globular and Membrane proteins?
- Are proteins random sequences of amino acids (a long discussion)?

For these questions, the answers are as follows:

- There exists significant syntactic difference between Globular and Membrane proteins, and this difference is described;
- Amino acid sequences in proteins are definitely not random.

## 1  INTRODUCTION

Protein is the main engineering material of living cells. The idea of "protein-machine" approach to living cells biophysics was suggested in 1967 (see the review written by the authors of this idea and devoted to its 20[th] birthday (Chernavskii et al., 1987)).

The kinetic approach to protein-machines was reported in (Kurzynski, 1998). The experimentally justified models of conformational transition dynamics are described and applied to the interpretation of a few simple biochemical processes.

The engineering approach to protein-machines based on energetic classification of elements was discussed in (Alberts, 1998). Engineers recognize certain fundamental behaviors in nature and then create an idealized element to represent each of those behaviors. The heart of such methods is the simplification and idealization of a real world machine as a composition of discrete elements. Any particular part of a machine might be modeled as consisting of one or more of these basic constituent elements.

We are in desperate need of a tool for automatic disassembling of protein machines onto *elementary functional details*. These functional details should have a lumped description that is maximally independent of the materials: a wheel is the wheel, and the range of material variations can be very wide.

The definition of a functional detail is predetermined by its functional properties: flexibility or rigidity, affinity, etc. These properties should be described for a given function, and the set of realizations of the described properties in material (amino acids, for example) should be prepared. This is a long honest way, a physical and chemical description of protein machines functional details. Our hypothesis is that a "King's way" is possible for this description. It is the *entropic (informational) classification* (Bugaenko et al., 1998b).

---

* To whom correspondence should be addressed.

Now, the huge amount of protein primary structures is known. We can assume that *non-random differences and similarities* between these structures are determined by theirs functions. We should just extract these typical differences and similarities, and prepare the collection (a warehouse in a form of a dictionary) of these details and its material realization.

In this paper we start from the amino acid alphabet reduction and try to find the rigorous statement and the exact answer to the question: in which alphabet the protein primary structure is the maximally non-random sequence. The following steps will consist in analysis of $q$-letter ($q = 2, 3, 4$) fragments of proteins written in reduced alphabet with the same question and further the $q$-letter fragments alphabet reduction to consider recoded protein again thus increasing words length and extracting functional details.

Amino acid classifications and alphabet reduction are widely used in various applications. For example, physical and chemical properties of amino acids provide a number of natural classifications (Kawashima et al., 2000) and the most popular of them hydrophobicity/polarity binary classification revealed many interesting patterns of secondary structure preferences (see recent (Hennetin et al., 2003, Phoenix et al., 2002) and rather old (Eisenberg et al., 1984, West et al., 1995) papers).

Recently the idea of amino acid alphabet reduction is developed in two directions. The first one is related to protein design and the main question is: How many amino acids are sufficient to protein folding and function in a correct way (see review (Chan, 1999)). In the second direction we unite various theoretical approaches for amino acid grouping based on substitution matrices, secondary structure, pattern conservation and so on. These methods aim at simplification of amino acid sequences in order to solve some problems of sequence analysis resulted from alphabet size, such as revealing distinct homologs, secondary structure prediction, folds recognition and so on. Unlike protein design approach amino acids are to be classified (not selected) to reduce alphabet size.

Theoretical groupings of amino acids mentioned in literature may be attributed to the following main bases: (1) physical, chemical properties (Sirabella et al., 1998) and amino acid environment in proteins (Liu et al., 2003); (2) protein alignments (Irback et al., 1997) and substitution matrices (Liu et al., 2002, Murphy et al., 2002, Li et al, 2003); (3) protein spatial structure (Chen et al., 2002, Solis et al., 2002) and contact potential matrix (Li et al., 1997, Wang et al., 1999, Wang et al., 2000, Wang et al., 2002, Fan et al., 2003, Cieplak et al., 2001); (4) information theory (Bugaenko et al., 1998, Mintseris et al., 2004, Solis et al., 2002). One should understand that pattern conservation, substitution matrices, spatial structures and contact potentials are dependent on physicochemical properties of amino acids, that is why, as it was shown in papers mentioned above, all amino acid classifications based on the famous amino acid related matrices such as BLOSSUM, PAM or MJ-matrix correlate with hydrophobicity/polarity scale.

In this paper we consider the method of informational classification of amino acids that have been proposed in (Bugaenko et al., 1998b) for classification of amino acids in protein sequences and recently it was used in (Mintseris et al., 2004) for classification of proteins heavy atoms on the basis of statistical contact potential matrix. The notion of *informational classification* relates to entropic optimality principle that may have various formulations while the main idea consists in selection such classification that maximizes nonrandomness. Method of informational classification has universal nature: it can reduce any alphabet on the basis of $q$-letter statistic of some language objects. For example, there was spatial contacts statistic in (Mintseris et al., 2004) or linear contacts statistic in (Bugaenko et al., 1998b).

Informational classifications based on entropic optimality criterion were mentioned in a number of papers while fundamental properties of result such as stability, existence and uniqueness of optimal classifications are still not considered at all. Further we present: (1) complete formulation of optimality principle, (2) application of the method to amino acid classification on the basis of protein sequences, (3) regular analysis of results obtained by applying the method to various groups of proteins, words lengths, and so on, (4) comparison of some reduced amino acid alphabets with informational ones.

## 2  INFORMATIONAL CLASSIFICATIONS

### 2.1  Entropic optimality principle

The problem of amino acid *classification* into $k$ groups may be formalized as follows: given a set of 20 amino acids and a number of groups $k$ ($1 < k < 20$) one should find a mapping $f$ of amino acids onto groups $\{A,C,D,E,F,G,H,I,K,L,M,N,P,Q,R,S,T,V,W,Y\} \xrightarrow{f} \{1, \ldots, k\}$ so that every amino acid is assigned to one group and some optimality condition holds.

The notion of *informational classification* relates to the entropic optimality condition. Similar to (Bugaenko et al., 1998b, Mintseris et al., 2004, Gorban et al, 2005) we consider amino acid mapping to be optimal if recorded objects have as informative distribution as possible. Here "informative" means far from random while random distribution can be determined in various ways. Using well-known notion of relative entropy the optimality condition looks as follows. *Optimal informational classification provides maximal relative entropy of distribution of recorded objects:*

$$f: \quad D(P \mid P^*) = \sum_{X_f} P(X_f) \ln \frac{P(X_f)}{P^*(X_f)} \to \max, \qquad (1)$$

where $X_f$ represents recoded objects, $P(X_f)$ is real distribution, and $P^*(X_f)$ is some reference ("random") distribution of $X_f$.

Relative entropy is non-negative and $D(P \mid P^*) = 0$ iff $P = P^*$ (that means coincidence of real and reference distributions.) $D(P \mid P^*)$ is the convex functional of $P$, and, therefore, has one zero minimum, but can have many local maxima. It is worth to mention that the relative entropy in (1) measures additional information in the distribution $P$ with respect to the distribution $P^*$. In other words, it is the indicator of the relative order in $P$ against a background of $P^*$. The reference distribution in (1) generalizes criteria introduced in (Bugaenko et al., 1998b, Mintseris et al., 2004, Gorban et al, 2005).

$X$ may represent any structural objects of corresponding language. While any classification $f$ determines new alphabet and therefore a set of recoded objects $X_f$ with distribution $P(X_f)$. If one consider protein sequences than $X$ are sequence fragments of some length $q$; if one consider protein spatial structure than $X$ are the set of amino acid contacts in folded protein (like in (Mintseris et al., 2004)) and so forth.

### 2.2  Informational classification of amino acids

Here we consider amino acid classification on the basis of protein primary structure so we use $q$-letter words distribution and reconstructed $q$-letter words distribution as reference one. Reconstructed distribution is calculated according to some model of sequence generation. For example the most popular reconstruction of frequencies of $q$-letter words ($q > 1$) is simple multiplication of all $q$ letter frequencies. Markov model can be used as well. However we kept in mind the idea of the maximum entropy principle for frequency reconstruction proposed in (Bugaenko et al., 1998a). According to it reconstructed $q$-letter words distribution should have maximum entropy among all $q$-letter words distributions that could be obtained from the given ($q$-$s$)-letter word distribution. Thus, one can get reconstructed theoretical $q$-letter words frequencies on the basis of any less length words distribution. Relative entropy of real distribution with respect to reconstructed one shows how much information is introduced by $q$-letter words compared to ($q$-$s$)-letter words (Gorban et al., 2001). Hence alphabet reduction that maximizes (1) makes the most nonrandom $q$-letter words distribution given ($q$-$s$)-letter words distribution.

In more detail, given some protein sequence or a set of sequences one can obtain a vector of frequencies (or frequency dictionary) of protein fragments of the length $q$ (denoted here $W_q$) According to some classification of amino acids into $k$ groups the $W_q$ can be easily transformed to vector of recoded words frequencies $RW_q$ by simple summation. Having chosen the type of reference distribution one can obtain reconstructed frequencies for the same length $q$ $R\widetilde{W}_q$. Relative entropy is calculated as follows:

$$D(RW_q \mid R\widetilde{W}_q) = \sum_{i_1,i_2,..i_q \in \{1..k\}} f_{i_1 i_2 ... i_q} \ln \frac{f_{i_1 i_2 ... i_q}}{\widetilde{f}_{i_1 i_2 ... i_q}}. \quad (2)$$

Optimal amino acids classification corresponds to maximal value of $D(RW_q \mid R\widetilde{W}_q)$ for a given number of classes $k$.

Reconstructed frequencies according to (Bugaenko et al., 1998a) are calculated as follows.

$$\widetilde{f}_{i_1...i_{q-s}...i_q} = \frac{f_{i_1...i_{q-s}} \cdot f_{i_2...i_{q-s+1}} \cdot ... \cdot f_{i_{s+1}...i_q}}{f_{i_2...i_{q-s}} \cdot f_{i_3...i_{q-s+1}} \cdot ... \cdot f_{i_{s+1}...i_{q-1}}}, \text{ for } q - s > 1$$

and (3)

$$\widetilde{f}_{i_1...i_q} = f_{i_1} \cdot f_{i_2} \cdot ... \cdot f_{i_q}, \text{ for } q - s = 1,$$

where $i_q$ are symbols of alphabet, $i_1...i_q$ are $q$-letter words and $f_{i_1...i_q}$ are frequencies of corresponding words in the symbol sequence.

Such formulation gives a set of reference distributions to use in (2) dependent on the words length $q$ and on the "depth" of reconstruction $s$. Thus optimal informational classification of alphabet symbols produces a recoded sequence with maximal nonrandomness in frequencies of pairs with respect to symbol frequencies or maximal nonrandomness in frequencies of triplets with respect to pairs and so on in any reasonable combination. In this paper we always use reconstruction on the basis of single letter frequencies (the case $q - s = 1$ in (3)), its worth to note that this calculation coincides with zero order Markov model (independent letter generation model).

### 2.3 Binary informational classifications

Consider the problem of informational binary classification of amino acids based on 2-letter word distribution calculated for some set of proteins. The problem of binary classification consists in separation of amino acids into two groups: {A,C,D,E,F,G,H,I,K,L,M,N,P,Q,R,S,T,V,W,Y}→{0,1}.

Initial vectors of frequencies $W_1$ and $W_2$ are calculated for the set of protein sequences and then they are recalculated to binary ones by simple summation according to current amino acid grouping:

$$W_1 = \{f_A, f_C, \ldots, f_Y\} \rightarrow RW_1 = \{f_0, f_1\}$$

$$W_2 = \{f_{AA}, f_{AC}, \ldots, f_{YY}\} \rightarrow RW_2 = \{f_{00}, f_{01}, f_{10}, f_{11}\}$$

Reconstructed frequencies for $q = 1$ are calculated as follows:

$$\widetilde{f}_{ij} = f_i \cdot f_j, \quad i, j = 0, 1,$$

$$R\widetilde{W}_2 = \{\widetilde{f}_{00}, \widetilde{f}_{01}, \widetilde{f}_{10}, \widetilde{f}_{11}\}.$$

Following some scheme of enumeration of amino acid groupings, one calculates the relative entropy

$$D(RW_2 \mid R\widetilde{W}_2) = \sum_{i_1, i_2 \in \{0,1\}} f_{i_1 i_2} \ln \frac{f_{i_1 i_2}}{\widetilde{f}_{i_1 i_2}} \quad (4)$$

for each grouping and chooses the most informative one.

We used the simplest binary problem in order to describe some important things about amino acid classification, such as existence and uniqueness of global maximum of (2), its stability dependent on frequency distribution and so on.

In order to achieve global maximum (2) an exhaustive search is needed. We used quasi-random optimization that can be specified as one-step directed descent with a number of random starts. The method consists in following steps: 1) choosing some random classification as a start point and calculating relative entropy (2), 2) calculating relative entropy for all one-letter neighbor classifications, 3) choosing amino acid grouping that produce maximal increment in relative entropy as the next start point, 4) going on to the step 2 until any increment in relative entropy exists. We repeat procedure with 100 random start points and choose the best classification. Method showed more than 99% hit to the global maximum (for 116 protein sequences method produced 115 optimal classifications).

## 3 DATASETS

The method of entropic classification may reduce any alphabet on the basis of some symbol sequence (or sequences) written in this alphabet. As far as amino acids concern, every single protein sequence could have its own optimal informational classification of amino acids that makes it at maximum non-random given word length and alphabet size. Any set of protein sequences could have its own classification of amino acids and so on. There exist some natural types of protein sets which can be considered as a basis of classification: protein families (superfamilies), proteins with similar function, structure or subcellular localization, proteomes, all known protein sequences as a whole and so on.

Table 1. The data sets of protein sequences

|  | Keywords | Numb. of proteins | Keywords | Numb. of proteins |
|---|---|---|---|---|
| **Dataset 1 (EBI)** | Oxidoreductase | 452 | Transferase | 500 |
|  | Cytochrome | 500 | Isomerase | 578 |
|  | Phytochrome | 500 | DNA polymerase | 500 |
|  | Nitratoreductase | 197 | Oxidase | 500 |
|  |  |  | ATPase | 500 |
| **Dataset 2 (SwissProt)** | Membrane | 10000 | Globular | 5019 |

We used several sets of proteins extracted by keyword from databases. Table 1 shows two datasets with the number of protein sequences extracted: Dataset 1 and Dataset 2 were formed according to the keyword in the field Definition from EBI (http://www.ebi.ac.uk) and Swissprot (http://expasy.org) databases correspondingly.

As one can see Dataset 1 was completed by sets of proteins with similar function and sets of Dataset 2 can be characterized to have similar structure and subcellular localization. These protein sets were natural choice to elaborate the properties of informational classifications of amino acids.

## 4 RESULTS AND DISCUSSION

### 4.1 Informational classification for large groups of proteins

Optimal informational **binary** classifications of amino acids were calculated for sets of proteins as follows. Vectors of frequencies of 2- and 1-letter words were counted through whole sets of proteins and optimal binary classifications were obtained (Tables 2).

**Table 2.** Binary informational classifications for Dataset 1 and 2

| Protein dataset | $D_{max}$ | A | C | D | E | F | G | H | I | K | L | M | N | P | Q | R | S | T | V | W | Y |
|---|---|---|---|---|---|---|---|---|---|---|---|---|---|---|---|---|---|---|---|---|---|
| ATPase | 0.002 | 1 | 1 | 0 | 0 | 1 | 1 | 1 | 1 | 0 | 1 | 1 | 0 | 0 | 0 | 0 | 1 | 1 | 1 | 1 | 1 |
| Cytochrome | 0.0066 | 1 | 1 | 0 | 0 | 1 | 1 | 1 | 1 | 0 | 1 | 1 | 0 | 0 | 1 | 0 | 1 | 1 | 1 | 1 | 1 |
| Nitratoreductase | 0.0027 | 1 | 0 | 0 | 0 | 1 | 1 | 1 | 1 | 0 | 1 | 1 | 0 | 0 | 0 | 0 | 1 | 1 | 1 | 1 | 1 |
| Oxidase | 0.0029 | 1 | 1 | 0 | 0 | 1 | 1 | 1 | 1 | 0 | 1 | 1 | 0 | 0 | 0 | 0 | 1 | 1 | 0 | 1 | 0 |
| DNA polymerase | 0.0007 | 1 | 0 | 0 | 1 | 0 | 0 | 0 | 0 | 1 | 0 | 1 | 0 | 0 | 0 | 1 | 0 | 0 | 0 | 0 | 0 |
| Isomerase | 0.0006 | 1 | 0 | 1 | 1 | 0 | 0 | 0 | 1 | 1 | 1 | 0 | 0 | 1 | 1 | 1 | 0 | 0 | 1 | 0 | 0 |
| Transferase | 0.0006 | 1 | 0 | 1 | 1 | 0 | 0 | 0 | 1 | 1 | 1 | 0 | 0 | 0 | 1 | 0 | 0 | 0 | 0 | 0 | 0 |
| Phytochrome | 0.0074 | 1 | 1 | 1 | 1 | 0 | 0 | 1 | 1 | 1 | 0 | 1 | 1 | 0 | 0 | 0 | 0 | 0 | 1 | 0 | 0 |
| Oxidoreductase | 0.0024 | 1 | 1 | 0 | 0 | 0 | 1 | 0 | 0 | 1 | 1 | 0 | 1 | 1 | 0 | 1 | 1 | 0 | 1 | 0 | 1 |
| Globular | 0.0006 | 1 | 0 | 0 | 1 | 0 | 0 | 0 | 0 | 1 | 1 | 1 | 0 | 0 | 1 | 1 | 0 | 0 | 0 | 0 | 0 |
| Membrane | 0.0025 | 1 | 1 | 0 | 0 | 1 | 1 | 0 | 1 | 0 | 1 | 1 | 0 | 1 | 0 | 0 | 1 | 1 | 1 | 0 | 0 |

As one can see, considered sets of proteins provide different {0, 1} amino acid informational classifications. However more careful consideration shows that there exist similar classifications as well as absolutely different ones. Moreover, the values of maximal relative entropy are close to zero. Zero minimum corresponds to equal real and calculated frequencies of 2-letter words. It means that even in optimal binary classification amino acid sequences are close to random ones if 2-letter words are considered. Our example with Bible (see section 4.4) shows that for English language it is not a case. Here binary classification provides significant value of relative entropy.

### 4.2 Properties of informational classifications of amino acids

*4.2.1 Distribution.* Analysis of distribution of binary classifications by relative entropy value shows that there exists one global maximum of relative entropy in the space of all possible binary classifications (Fig.1). The distribution is close to exponential and there are about 10 (as a rule similar) classifications in 90-100% interval of maximal relative entropy for all tested datasets.

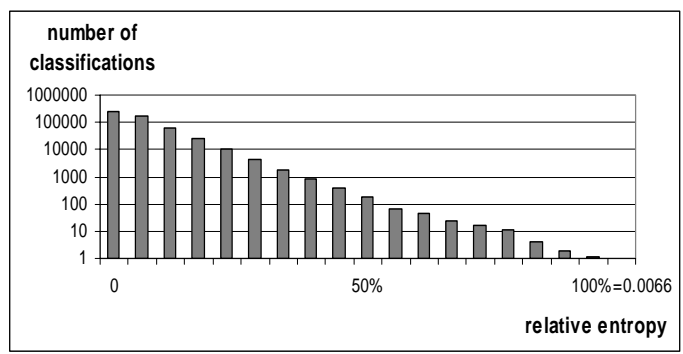

**Fig. 1.** Distribution of relative entropy value for all possible binary classifications of amino acids for Cytochrome dataset

*4.2.2 Correlation structure.* Correlation structure of binary classifications distribution by relative entropy value can become clear by constructing the system of successive "orthogonal" binary classifications.

The first optimal classification is chosen to provide maximum of functional (4). The second classification corresponds to constrained maximum of (4) with constrains introduced by the first one and so on. Here we considered two possible types of constraints imposed on new binary classification in hierarchy: 1) it should have Hamming distance equal to 10 to any previous one, and 2) it should have zero Pearson correlation with any previous one (actually, $|r| < 0.1$).

We kept in mind the idea of orthogonality and independency while introducing these constraints. As one can see the system of "orthogonal" classifications can be considered as an analog of principal components method in the problem of depletion of information in symbol sequences: the first classification produce maximal nonrandomness in words distribution, the second one should be independent of the first one and produce maximal nonrandomness and so on.

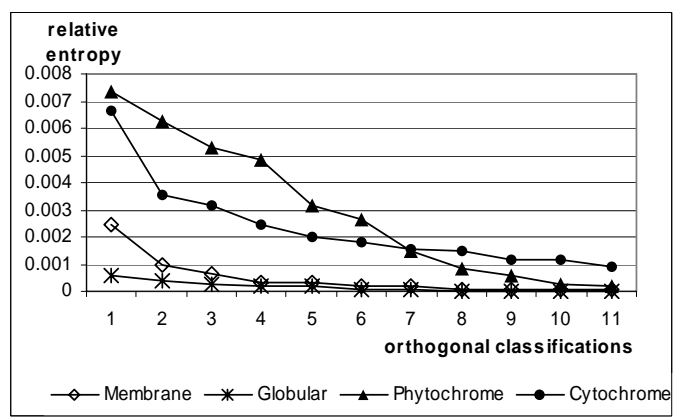

**Fig.2.** Relative entropy decrease for hierarchical orthogonal classifications for some datasets with Hamming distance constraints.

The graph of the relative entropy decrease (Fig.5) shows correlation structure of distributions. For example, one (Cytochrome dataset) and more than two (Phytochrome dataset) independent classifications are present in the top of relative entropy value.

*4.2.3 Classification stability.* Informational classifications seem to be very sensible to marginal changes in words frequencies. Amino acid frequency profiles seem rather similar for protein sets of Dataset 1 and 2 shown in Fig.3. The same situation is for 2-letter words dis-tribution. However, optimal classifications vary from similar to absolutely different ones. This kind of instability is rather natural property of optimization in the discrete space.

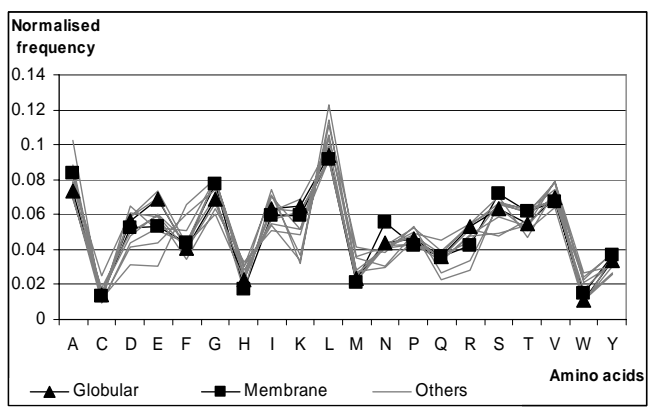

**Fig. 3.** Amino acid frequencies in considered sets of proteins

From the other side similarity of frequencies should provide high relative entropy of {0,1} recorded protein sequences following any optimal classification. However, it is not always a case. Relative entropy value as well can vary significantly: for some dataset there exists more than 10 times difference in relative entropy (see table 1). And, for example, Phytochrome and Oxidoreductase dataset in any other optimal classification display poor informativity (in average 3.5% and 5% of their maximal value).

*4.2.4 q = 2,3,4,5.* Binary informational classifications of amino acids on the basis of 3,4,5-letter word frequencies are obtained in the same way as for $q = 2$. The reference distribution is calculated according to (3) for $q - s = 1$, i.e. word frequencies are reconstructed on the basis of amino acid frequencies.

Five protein sets from datasets 1 have equal classifications for all basis word lengths. Other datasets have few equal classifications among the 4 length compared. It means that informational classification that maximizes nonrandomness in two letter words distribution produces maximal or near maximal nonrandomness in 3,4,5-letter words distribution.

The {0,1} words frequency profile of recorded datasets appeared similar too. For example, among all 16 4-letter words there is one with prevailed frequency for all datasets: this is 1111 word. It means that maximally non random {0,1} recoded proteins have the structure of repeated homogeneous series. This fact looks promising for revealing common structural elements in protein sequences.

*4.2.4 Correlation with some natural classifications.* There exists many well known physicochemical properties of amino acids and each of them may serve as a basis of amino acid classification. Here we consider five ones with three various versions of hydrophobicity/polarity (HP), big/small (BS), and charged/uncharged (CU) classifications (Table 3).

**Table 3.** Some natural amino acids binary classifications

| Property | A | C | D | E | F | G | H | I | K | L | M | N | P | Q | R | S | T | V | W | Y | Source |
|---|---|---|---|---|---|---|---|---|---|---|---|---|---|---|---|---|---|---|---|---|---|
| HP I   | 1 | 1 | 0 | 0 | 1 | 1 | 0 | 1 | 0 | 1 | 1 | 0 | 1 | 0 | 0 | 0 | 0/1 | 1 | 1 | 0/1 | 1 |
| HP II  | 1 | 1 | 0 | 0 | 1 | 1 | 0 | 1 | 0 | 1 | 1 | 0 | 0 | 0 | 0 | 0 | 0 | 1 | 1 | 1 | 2 |
| HP III | 1 | 1 | 0 | 0 | 1 | 0 | 1 | 1 | 0 | 1 | 1 | 0 | 0 | 0 | 0 | 0 | 0 | 1 | 1 | 1 | 3 |
| BS     | 0 | 0 | 0 | 1 | 1 | 0 | 1 | 1 | 1 | 1 | 1 | 0 | 0 | 1 | 1 | 0 | 0 | 0 | 1 | 1 | 1 |
| CU     | 0 | 0 | 1 | 1 | 0 | 0 | 1 | 0 | 1 | 0 | 0 | 0 | 0 | 0 | 1 | 0 | 0 | 0 | 0 | 0 | 1 |

Source: 1. http://en.wikipedia.org/wiki/Amino_acid; 2. (Eisenberg et al.,,1984); 3. (Cid et al., 1992)

Direct comparison of optimal informational binary classifications and natural binary classifications shows similarity between Hydrophobic / Polar classification and informational ones for ATPase, Cytochrome, Nitratoreductase, Oxidase and Membrane datasets (r ≈ 0.7). The same classifications have moderate similarity with Charged / Uncharged binary pattern (r ≈ 0.5). The detailed results for Dataset 2 are shown in Table 6.

### 4.3 Globular vs Membrane comparison

Consider informational classification for Globular (G) and Membrane (M) datasets in more details. We got two amino acid groupings:
G: {A,**E,K**,L,M,**Q,R**} and {**C**,D,**F,G**,H,**I**,N,**P,S,T,V**,W,Y},
M: {D,**E**,H,**K**,N,**Q,R**,W,Y} and {A,**C,F,G,I**,L,M,**P,S,T,V**}.

Hamming distance for these classifications is equal to 8. It means that they are practically independent (orthogonal) Amino acid frequency profiles for these datasets are similar (fig. 3). However, classifications are rather different. Membrane binary classification is nicely correlated with Hydrophbic/Polar I and Charged/Uncharged classification (r=0.7, and 0.63, Hamming distance=2 and 4). Globular classification has moderate correlation with Big/Small property (r=0.45, Hamming distance=6). So, approximate coincidence of Hydrophbic/Polar and informational classification for Membrane dataset shows principal importance of that feature for these proteins structural organization. This is not surprising. But it seems more important that for globular proteins the main amino acid classification is very specific and has almost no similarity to usual classifications (only the size of amino acid matters slightly).

### 4.4 En the netennent ten the Tent, ent the Tent ten teth Tet, ent the Tent ten Tet

Method of informational classification is rather universal: having some text one can reduce its alphabet and the text written in reduced alphabet would be at most non-random. For example, we consider the text of King James Bible http://bible.cc written in English language. 2-letter frequencies were calculated over the whole words. Gaps and numbers were ignored. Reference frequencies were calculated as product of symbol frequencies. Optimal classifications of alphabet symbols in 2-6 groups are shown in Table 4.

**Table 4.** Entropic classification of letters for English language in Bible text

| Relative Entropy | Groups | | | | | |
|---|---|---|---|---|---|---|
| | 1 | 2 | 3 | 4 | 5 | 6 |
| 0.767926 | a**e**ioudgt | Bcfhjklm**n**pqrsvwxyz | | | | |
| 0.934107 | a**e**iou | Bcdfgklmnpqrs**t**vwxyz | **h**j | | | |
| 1.096432 | a**e**iou | Bcfklm**n**pqrsvxz | **h**j | dg**t**wy | | |
| 1.171895 | a**e**iou | Bcfklmpr**s**vxyz | **h**j | dgq**t**w | **n** | |
| 1.227138 | a**e**iou | Bcfklmpqr**s**vxyz | **h**j | **t**w | **n** | **d**g |

The most frequent letter is written in bold underlined according to standard frequency table of English language letters (Lewand et al., 2002).

Surprisingly, letters that correspond to a vowel appeared to constitute the group. It means that recoding text according to vowel and consonant one gets very non-random text. The informational classification "discovered" the syllabic structure of the text (with some approximation error). It is very natural. For example, in the optimal binary classification the text should be maximally ordered locally. It means either the presence of long one-class clusters (0000… or 11111…), or local order of the form 0101… The first possibility cannot be realized for the natural language. Hence, for the best binary classification the encoded texts should be locally close to 0101… . And it is the syllabic structure (approximate syllabic structure, to be exact).

And now we can explain the title of the section: "*In the beginning was the Word, and the Word was with God, and the Word was God.* The same was in the beginning with God. All things were made by him" (Jn. 1:1-3).

**Table 5.** The phrase coded by the most frequent letters from classes (with an example of {01} coding)

| Number of classes | The recoded phrase |
|---|---|
| 2 | `01 110 101011011 101 110 1011 011 110 1011 101 1011 101  011 110 1011 101 101`<br>`En nne nenennenn nen nne Nenn, enn nne Nenn nen nenn Nen, enn nne Nenn nen Nen` |
| 3 | `Et the tetettett tet the Teth, eth the Teth tet teth Teh, eth the Teth tet Teh` |
| 4 | `En the netenent ten the Tent, ent the Tent ten teth Tet, ent the Tent ten Tet` |
| 5 | `En the setenent tes the Test, ent the Test tes teth Set, ent the Test tes Set` |
| 6 | `En the setennend tes the Tesd, end the Tesd tes teth Ded, end the Tesd tes Ded` |
| Initial phrase | `In the beginning was the Word, and the Word was with God, and the Word was God` |

## 5  CONCLUSION AND OUTLOOK

We presented the method of informational classification of amino acids based on protein sequences and considered some interesting properties of informational binary classifications. Informational binary classification makes protein sequence at maximum non random. However, relative entropy of recoded protein sequences is close to zero and it is not a case for Bible. So, even in optimal binary alphabet protein sequences slightly differ from zero order model.

We revealed that informational classification that maximizes nonrandomness in two letter words distribution produces maximal or near maximal nonrandomness in 3,4,5-letter words distribution. We found similar 4-letter binary words frequency profile for all datasets, so, maximally non random {0,1} recoded proteins have the structure of repeated homogeneous series. This fact looks promising for revealing common structural elements in protein sequences.

We showed the existence and uniqueness of optimal classification. We observed correlation between informational classifications and natural classifications of amino acids such as Hydrophobic/Polar and Charged/Uncharged ones for some datasets.

Informational binary classifications are sensitive to amino acid distribution. However, similarity of one and two letter words frequency profiles in a number of considered protein sets provides as a rule high relative entropy for protein sequences recoded in any optimal classification.

**Table 6.** Hamming distances between various binary classifications

|  | HP I | HP II | HP III | BSl | CU | Membr. | Glob. | Membr. +Globr |
|---|---|---|---|---|---|---|---|---|
| Membrane | 2 | 5 | 7 | 6 | 4 | 0 | 8 | 6 |
| Globular | 7 | 9 | 9 | 6 | 6 | 8 | 0 | 8 |
| Membrane+Globular | 5 | 7 | 7 | 10 | 4 | 6 | 8 | 0 |
| (Cieplak et al., 2001)* | 3 | 4 | 2 | 7 | 7 | 7 | 7 | 7 |
| (Irback et al., 1997) | 3 | 2 | 2 | 7 | 7 | 7 | 9 | 9 |
| (Murphy et al., 2002) | 3 | 3 | 5 | 8 | 2 | 2 | 6 | 4 |
| (Liu et al., 2002) MJ | 3 | 3 | 3 | 8 | 8 | 6 | 8 | 10 |
| (Liu et al., 2002) BLOSSUM | 4 | 3 | 3 | 6 | 8 | 8 | 10 | 10 |

*(Chen et al.,2002) and (Li et al, 2003) have the same amino acid groupings.

Finally, in Table 6 we present a pair comparison of amino acid binary classifications obtained from three sources: 1) informational classifications for Membrane, Globular dataset and Dataset 2 as a whole; 2) some natural classifications; 3) classifications taken from recent publications. Amino acid groupings mentioned in reviewed papers do have similarity with Hydrophobic/Polar classification while some informational classifications show similarity to Hydrophobic/Polar and Charged/Uncharged property. Classification of (Murphy et al., 2002) is the only one to be rather close to informational classifications.

The idea of amino acids to be structural units of the sequence which contain all the information to protein fold and function is the base metaphor in studying amino acids informational classifications. The simplest binary informational classification gives the most nonrandom recoded sequences which allows one to study words distribution and to extract functional units.

Fundamental property of informational classification method is its universal nature: the method may reduce any alphabet on the basis of frequencies of words of any length. Thus one can consider other structural units which probably determine protein fold and function. For example, the 400-letter alphabet of amino acid pairs can be supposed to be structural units. And in this case an alphabet reduction is the key operation. Presented method can perform alphabet reduction to be formal, automatic and the most informative.

Another possible application of the method is amino acid classification based on contacts statistics. To the moment only pair contacts are considered while the method allows using an arbitrary set of contacts. Informational classification makes recoded sets of contacts to be at the most nonrandom.

The nearest steps in informational classification method development and application could be as follows:
(1) Computation of optimal 2, 3, 4, and 5-letters alphabets for databases of known protein primary structures;
(2) The informational classifications analysis of 2 and 3-letters elements of primary structures (after reduction of the initial alphabet to 2-5 symbols) and so on, producing hierarchy of constructive details.

How can we answer now the main question: What are proteins made from? Are they made from the following details:
- Amino Acids (AAs)?
- Short sequences of AAs?
- Classes of equivalent AAs?
- Short sequences of such classes?
- Anything else?

For the set of available globular proteins we find the following main binary classification of AAs:

G: {A,E,K,L,M,Q,R}U{C,D,F,G,H,I,N,P,S,T,V,W,Y}.

For the set of available membrane proteins produces the following main binary classification which almost coincides with hydrophobic/hydrophilic classification:

M: {D,E,H,K,N,Q,R,W,Y}U{A,C,F,G,I,L,M,P,S, T,V}.

The primary sequences of proteins for these classes are definitely not random, and this non-randomness (the correlations between AAs in sequences) is significantly different for these classes. If we combine two classifications into one "GorM" classification, then we get four groups

{$\underline{A}$,**$\underline{L}$**,M}U{C,F,**$\underline{G}$**,I,P,$\underline{S}$,T,V}U{$\underline{E}$,**$\underline{K}$**,Q,R}U{**$\underline{D}$**,H,$\underline{N}$,W,Y}.

In each class, we select two most frequent AAs. The symbol of the most frequent AA is written in **bold underlined**. The symbol of the second AA is written in *italic underlined*. Using these AAs we can call the classes as L-A (Leucine-Alanin) group {$\underline{A}$,**$\underline{L}$**,M}, G-S (Glycine-Serine) group {C,F,**$\underline{G}$**,I,P,$\underline{S}$,T,V}, K-E (Lysine-Glutamic Acid) group {$\underline{E}$,**$\underline{K}$**,Q,R}, and D-N (Aspartic Acid-Asparagin) group {**$\underline{D}$**,H,$\underline{N}$,W,Y}.

So, we have new candidates for a minimal set of amino acids (one AA from a group or two AAs from a group). But, perhaps, it is wiser to classify couples and triples of amino acids. Classes of such couples and triples are, perhaps, the **elementary** details of proteins.

**ACKNOWLEDGEMENTS.** We are grateful to M. Gromov for the question about difference between informational classifications for Membrane and Globular proteins.